\documentclass[pdflatex,sn-aps]{sn-jnl}

\usepackage[normalem]{ulem}
\usepackage{caption}
\usepackage{graphicx}%
\usepackage{multirow}%
\usepackage{amsmath,amssymb,amsfonts}%
\usepackage{amsthm}%
\usepackage{mathrsfs}%
\usepackage[title]{appendix}%
\usepackage{xcolor}%
\usepackage{textcomp}%
\usepackage{manyfoot}%
\usepackage{booktabs}%
\usepackage{algorithm}%
\usepackage{algorithmicx}%
\usepackage{algpseudocode}%
\usepackage{listings}%

\theoremstyle{thmstyleone}%

\theoremstyle{thmstyletwo}%

\theoremstyle{thmstylethree}%

\begin{document}

\title[Article Title]{Entangled dual-comb spectroscopy}

\author[1]{\fnm{Abdulkarim} \sur{Hariri}}

\author[1]{\fnm{Shuai} \sur{Liu}}

\author[2]{\fnm{Haowei} \sur{Shi}}

\author[2,3]{\fnm{Quntao} \sur{Zhuang}}

\author[4]{\fnm{Xudong} \sur{Fan}}

\author*[1]{\fnm{Zheshen} \sur{Zhang}}\email{zszh@umich.edu}

\affil[1]{\orgdiv{Department of Electrical Engineering and Computer Science}, \orgname{University of Michigan}, \orgaddress{\city{Ann Arbor}, \state{MI} \postcode{48109}, \country{USA}}}

\affil[2]{\orgdiv{Ming Hsieh Department of Electrical and Computer Engineering}, \orgname{University of Southern California}, \orgaddress{\city{Los Angeles}, \state{CA} \postcode{90089}, \country{USA}}}

\affil[3]{\orgdiv{Department of Physics and Astronomy}, \orgname{University of Southern California}, \orgaddress{\city{Los Angeles}, \state{CA} \postcode{90089}, \country{USA}}}

\affil[4]{\orgdiv{Department of Biomedical Engineering}, \orgname{University of Michigan}, \orgaddress{\city{Ann Arbor}, \state{MI} \postcode{48109}, \country{USA}}}

\abstract{Optical frequency combs have emerged as a cornerstone for a wide range of areas, including spectroscopy, ranging, optical clocks, time and frequency transfer, waveform synthesis, and communications. However, quantum mechanical fluctuations of the optical carrier impose fundamental performance limits on the precision of traditional classical laser frequency combs, particularly in their use for interferometry and spectroscopy. Entanglement, as a quintessential quantum resource, allows for surpassing the fundamental limits of classical systems. Here, we introduce and experimentally demonstrate entangled dual-comb spectroscopy (EDCS) that surmounts the fundamental limits of classical DCS. EDCS builds on tailored entangled spectral structures of the frequency combs, enabling simultaneous detection of all comb lines below the standard quantum limit of classical DCS. Applying EDCS in gas detection, we achieve a 2.6 dB enhancement in signal-to-noise ratio and a 1.7-fold reduction in integration time over classical DCS, rendering EDCS particularly suited for dynamic chemical and biological sensing, where fast, precise measurements subject to power constraints are required. EDCS represents a new paradigm for quantum frequency combs, underscoring their prospects in a plethora of applications in precision metrology, spectroscopy, and timekeeping.}

\maketitle

\section{Introduction}

Entanglement is a quantum-mechanical property with no classical equivalent, exhibiting stronger-than-classical correlations shared by two or more objects. Since its initial discovery~\cite{einstein1935can,schrodinger1935discussion,bell1964einstein}, entanglement has evolved from a subject for scientific debate to a transformative resource that enables capabilities beyond the limits of classical physics \cite{zhang2024entanglement}. Entanglement now a cornerstone of a broad range of fields, including quantum computing \cite{jozsa2003role}, quantum communication \cite{ekert1991quantum,ralph1999continuous}, quantum sensing \cite{polzik1992spectroscopy,xia2020demonstration,guo2020distributed,hao2022demonstration}, and quantum imaging \cite{pittman1995optical}, offers a pathway to surpass the fundamental limits of classical technologies arising from quantum mechanical fluctuations~\cite{zhang2024entanglement}. 

Optical frequency combs—a transformative tool comprising a spectrum of evenly spaced, phase-coherent laser lines—deliver exceptional performance in applications such as spectroscopy \cite{picque2019frequency, coddington2016dual}, ranging \cite{caldwell2022time}, optical clocks \cite{takamoto2005optical}, and waveform synthesis \cite{jiang2007optical}. In this context, dual-comb spectroscopy (DCS) has emerged as a powerful technique, renowned for its unmatched sensitivity and broad bandwidth within a short measurement time. Nevertheless, DCS exemplifies a classical approach constrained by a fundamental trade-off between sensitivity and measurement time imposed by quantum noise. With recent DCS demonstrations exhibiting quantum-noise-limited  capabilities \cite{xu2024near,walsh2023mode,camenzind2024shot}, a pivotal question arises: how can quantum resources be leveraged within a DCS configuration, as achieved in gravitational-wave detection~\cite{tse2019quantum}, distributed quantum sensing~\cite{xia2021quantum,xia2023entanglement}, and quantum illumination~\cite{zhang2015entanglement}, to exceed the such a fundamental limit?

Recently, protocols for quantum-enhanced spectroscopy~\cite{belsley2023quantum,shi2023entanglement} have emerged to overcome the fundamental limits of spectroscopy based on classical frequency combs. Specifically, Ref.~\cite{belsley2023quantum} proposed a scheme that leverages broadband squeezed light, electro-optic (EO) modulation, and homodyne detection to create a quantum frequency comb that would overcome the shot-noise limit of classical spectroscopy. However, the practical bandwidth of optical measurements precludes the proposed scheme from achieving a wide spectral coverage on par with that of the DCS. In parallel, Ref~\cite{shi2023entanglement} devised a DCS protocol based on a quantum comb comprising squeezed individual lines detected by a local oscillator (LO) comb in a heterodyne configuration. Unfortunately, the generation of the desired quantum comb would be a formidable task as it requires a nonlinear optical process with a phase-matching bandwidth narrower than the line spacing and a group-velocity-matching bandwidth as broad as that of the combs~\cite{shi2023entanglement}. As such, while these schemes yield theoretically promising results, the experimental realization of a metrological advantage in quantum frequency-comb spectroscopy remains an outstanding challenge.

Here, we introduce and experimentally demonstrate entangled dual-comb spectroscopy (EDCS) that overcomes the fundamental limits of classical DCS. By harnessing the entangled spectral structure across a large number of optical modes produced concurrently in a spontaneous parametric down-conversion (SPDC) process, EDCS enables simultaneous detection of all comb lines below the standard quantum limit (SQL), with a demonstrated signal-to-noise ratio (SNR) enhancement of 2.6 dB compared to classical DCS and a reduction in integration time by a factor of 1.7. Our experiment establishes a robust framework for integrating quantum resources into DCS, paving the way for transformative advancements in high-precision sensing and metrology.

\section{Principle of entangled dual-comb spectroscopy}\label{sec2}

Our approach to EDCS requires the use of three frequency-comb sources: an entangled comb to provide the quantum correlations needed to achieve sub-SQL performance; a classical comb that coherently displaces the entangled comb to produce a bright entangled signal comb transmitted to the sample; and a strong LO comb to simultaneously measure all entangled signal comb lines.

\begin{figure}[h]
\centering
  \includegraphics[width=\textwidth]{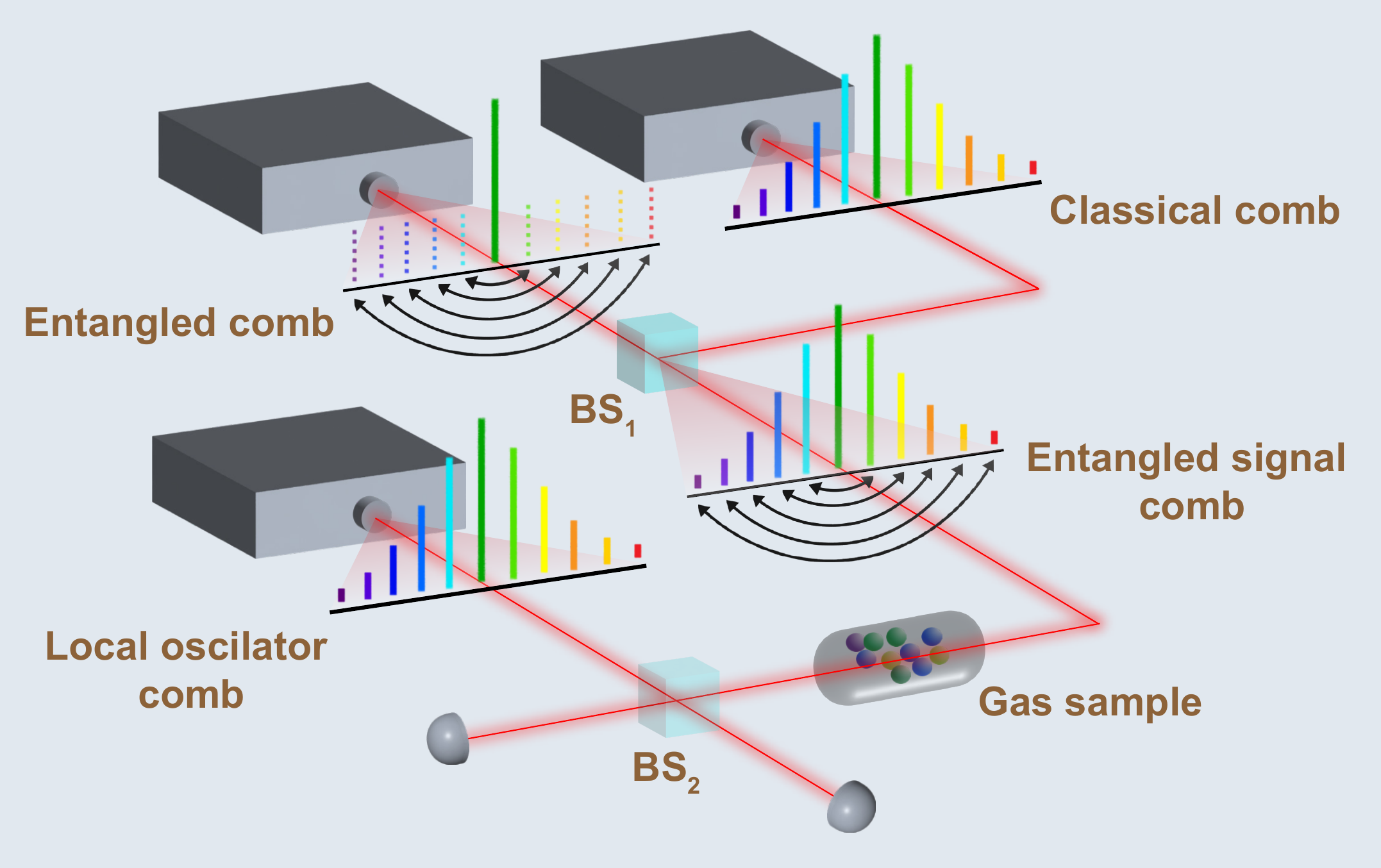}
    \caption{Experimental setup for entangled dual-comb spectroscopy. An entangled optical frequency comb is coherently displaced by a classical comb to generate a bright entangled signal comb. This signal comb is transmitted through a gas sample for spectroscopic analysis. The absorption fingerprint of the gas is retrieved through an interferometric measurement with a local oscillator comb in a heterodyne detection configuration. The protocol exploits the sub-SQL capabilities of the entangled comb to achieve a higher sensitivity and a reduced integration time compared to classical dual-comb spectroscopy. Solid arrows: entanglement shared by the comb lines. Dotted lines: two-mode squeezed vacuum states.}
    \label{fig:scheme}
\end{figure}

\begin{figure}[h]
\centering
  \includegraphics[width=\textwidth]{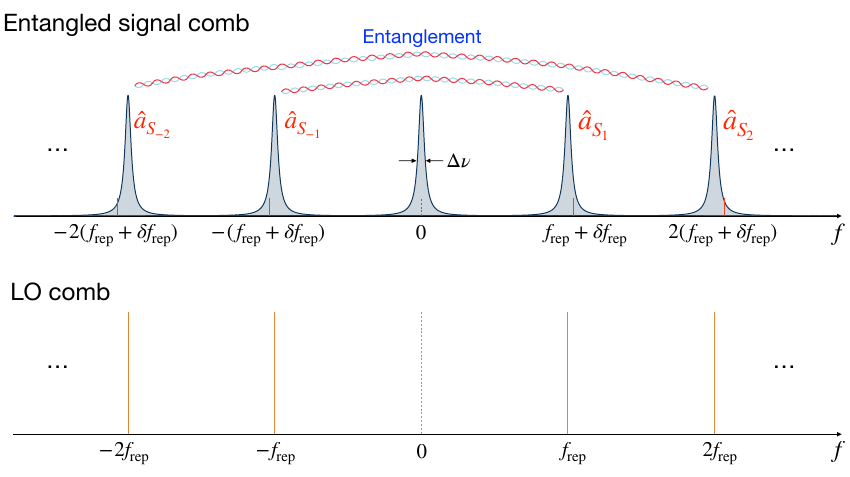}
    \caption{Entanglement structure and spectral configuration of entangled signal and LO combs. $\hat a_{S_n}$ and $\hat a_{S_{-n}}$ form pairs of TMSV state produced from the OPO cavity, constituting an entangled comb. The line spacing of the LO comb is locked to that of the entangled comb at $f_{\rm rep}$, allowing for simultaneous detection of all entangled signal comb lines below the SQL. The classical comb displaces the entangled comb by $n\alpha_{S_n}$ at frequencies $n(f_{\rm rep}+\delta f_{\rm rep})$ to produce a bright entangled signal comb. The central dashed comb lines are employed solely for phase locking.}
    \label{fig:EDCS_structure}
\end{figure}

The experimental schematic of EDCS shown in Fig.~\ref{fig:scheme} illustrates the interplay between the three comb sources. The quantum source based on a seeded optical parametric oscillator (OPO) cavity generates an entangled comb~\cite{chen2014experimental} described by the following Hamiltonian in the interaction picture~\cite{walls2011quantum}:
\begin{equation}
    \hat{H} = i\hbar \chi \sum_{n = 0}^N \left(\hat a_{S_{-n}}^\dag \hat a_{S_{n}}^\dag - \hat a_{S_{-n}} \hat a_{S_{n}}\right),
\end{equation}
where $\hat a_{S_n}$ and $\hat a_{S_n}^\dag$ are the annihilation and creation operators of the $n$th mode of the entangled comb, $\hbar$ is the reduced planck constant, and $\chi$ quantifies the strengths of the nonlinear interaction. Here, $\hat a_{S_{-n}}^\dag \hat a_{S_{n}}^\dag$ ($\hat a_{S_{-n}} \hat a_{S_{n}}$) describes the creation (annihilation) of a pair of photons in the $n$th and $-n$th modes through SPDC. All the mode pairs undergo parallel SPDC processes to produce an entangled comb with a spectral structure comprising a central comb line in a displaced single-mode squeezed state, while the remaining comb lines form pairs of two-mode squeezed vacuum (TMSV) state as depicted in Fig.~\ref{fig:EDCS_structure}. The total number of entangled comb lines, $N$, is determined by the phase-matching bandwidth of the SPDC over the free spectral range of the OPO cavity. To produce a bright entangled signal comb employed to interrogate the sample, the entangled comb interferes with a spectrally matched classical comb on an unbalanced beam splitter, converting each TMSV pair into a displaced two-mode squeezed state. After probing the sample, the entangled signal comb is measured in a heterodyne configuration, where it beats with a LO comb with a slight line-spacing offset to generate a radio-frequency (RF) comb acquired and processed by electronics. Both the LO and signal combs are produced by EO modulation on a continuous-wave (CW) laser followed by a programmable optical filter that configures the intensity and phase of each comb line (see Appendix). 

The EDCS protocol is benchmarked against the SQL defined by a classical quantum-noise-limited DCS protocol based on the same classical and LO comb configuration while the entangled comb is turned off. 

The entanglement structure embedded in the entangled signal comb and its spectral configuration along with that of the LO comb are illustrated in Fig.~\ref{fig:EDCS_structure}. The OPO cavity generates a single-mode squeezed state at the center and an entangled comb comprising pairs of TMSV state, as represented by the connected dotted lines and quantum operators $\hat a_{S_n}$ and $\hat a_{S_{-n}}$, that are multiples of $f_{\rm rep}$ away from the central cavity resonance, where $f_{\rm rep}$ is determined by the free spectral range of the OPO cavity. The linewidth, $\Delta \nu$, of the OPO cavity is sketched with the Lorentzian shape. The classical comb's line spacing differs from that of the entangled comb by $\delta f_{\rm rep}$ such that the interference between the two combs on  BS$_1$ leads to displacements $\alpha_{S_n}$ at frequencies $n(f_{\rm rep}+\delta f_{\rm rep})$, as represented by the red lines. The LO comb's line spacing is aligned with that of the entangled comb at $f_{\rm rep}$, allowing all entangled signal comb lines to be simultaneously measured at a quantum fluctuation level below that of the SQL by virtue of the property of the entanglement embedded in the TMSV states. At the same time, the beating between the LO comb line at $n f_{\rm rep}$ and the displacement on the entangled signal comb at $n(f_{\rm rep}+\delta f_{\rm rep})$ translates to a beat note at $n\delta f_{\rm rep}$ in the RF spectrum. The aliasing arising from the beating of the $\pm n$th lines of the entangled signal and LO combs is resolved by appropriately designing the displacement at each line (see Discussion section).

\begin{figure*}[h]
\centering
  \includegraphics[width=\textwidth]{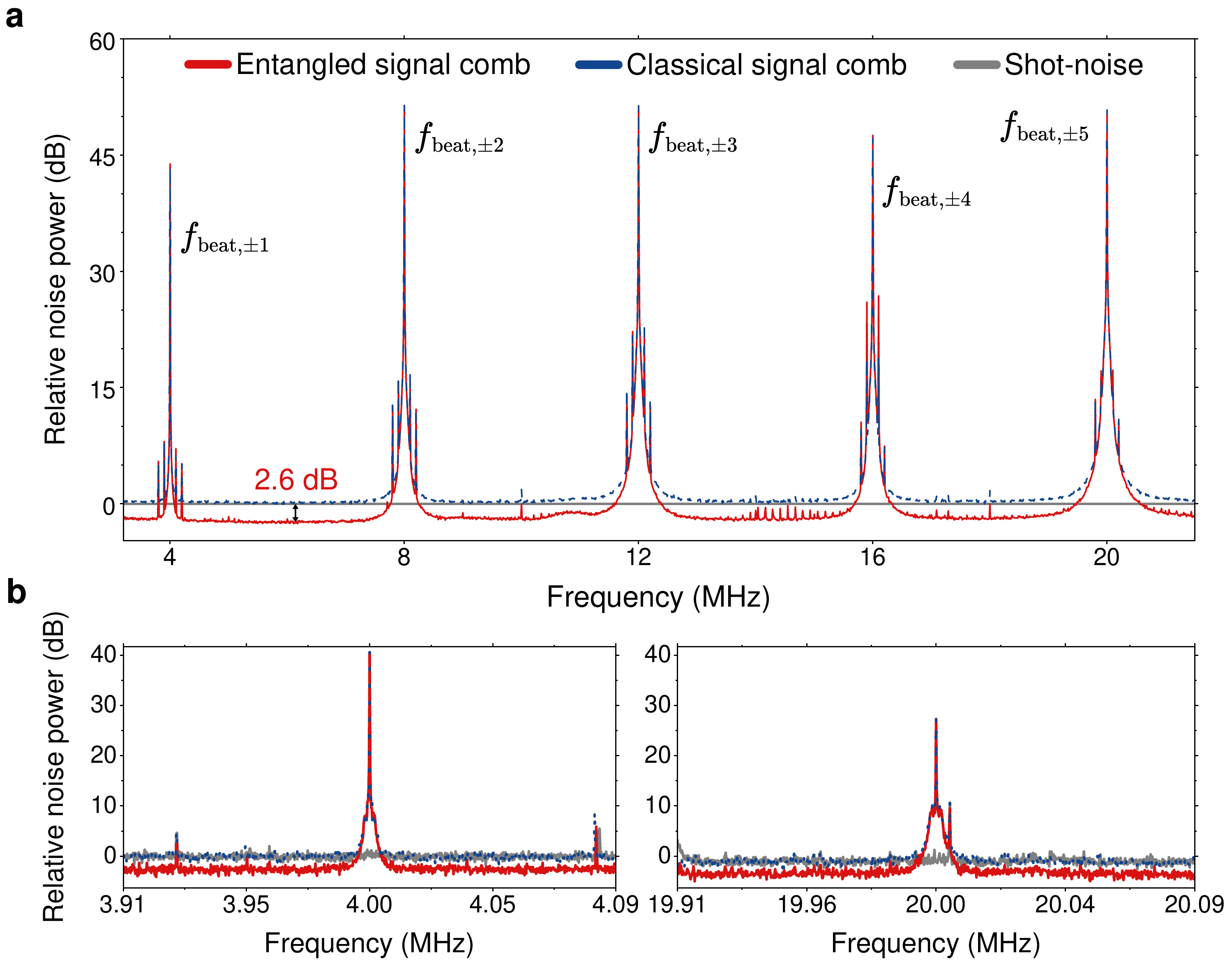}
    \caption{RF spectra in EDCS. (a) RF spectrum obtained from the beating of the entangled signal comb or classical signal comb with the local oscillator comb. Zoomed-in RF spectra at the 1st (b left panel) and 5th (b right panel) beatnotes obtained under reduced power levels of the classical signal comb. Operating at lower power mitigates the phase noise on the recorded beatnotes.}
    \label{fig:RF spectrum}
\end{figure*}

The RF spectrum resulting from the beating of the entangled signal or classical signal comb with the LO comb is depicted in Fig.~\ref{fig:RF spectrum}a, demonstrating enhanced SNR of 2.6 dB endowed by the entangled comb structure. To account for the increased phase noise at higher beat notes, the power of the classical signal comb is deliberately reduced. This adjustment ensures an expanded bandwidth for shot-noise-limited measurements, as illustrated in Fig.~\ref{fig:RF spectrum}b and Fig.~\ref{fig:RF spectrum}c.

\begin{figure*}[h]
\centering
  \includegraphics[width=\textwidth]{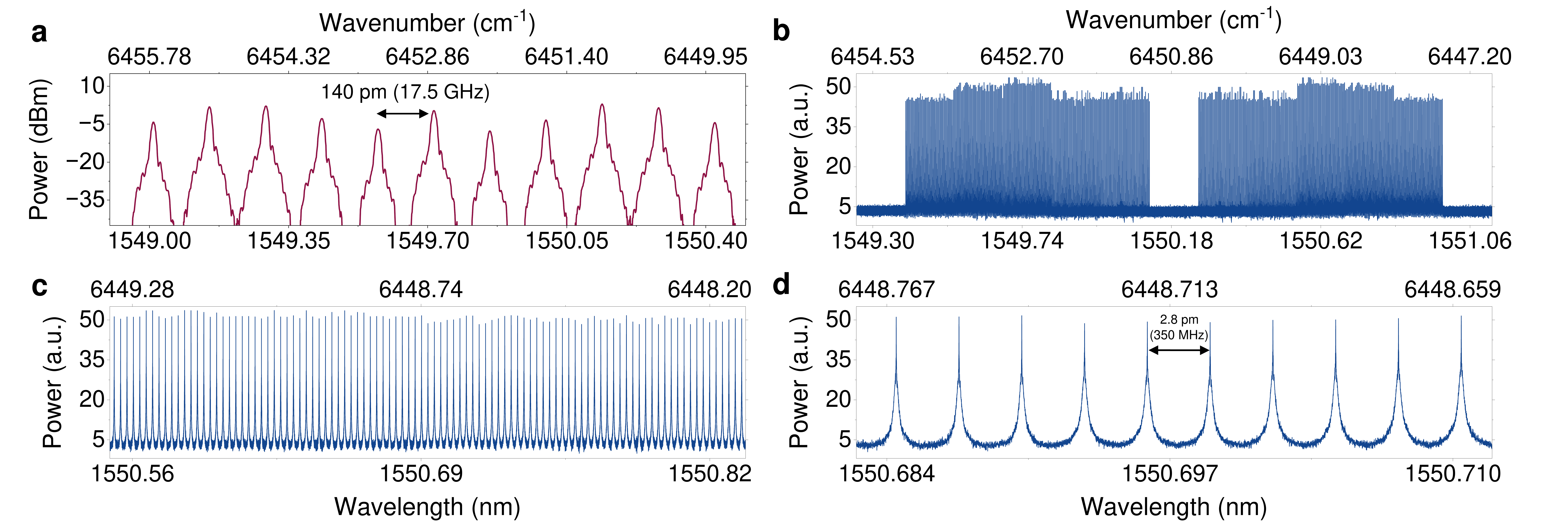}
    \caption{Measured local oscillator and dual-comb spectra. (a) Spectrum of the local oscillator comb measured using an optical spectrum analyzer. (b) Dual-comb spectrum obtained by sampling the time-domain signal of the subtracted photocurrents from the balanced receiver at a 100 MHz sampling rate for 0.5 seconds. The data is post-processed to derive the frequency-domain spectrum. The larger number of comb lines is achieved by sweeping the center frequency of the CW laser. (c) A zoomed-in view of a 35 GHz bandwidth, highlighting the mode-resolved nature of the dual-comb spectrum. (d) A further zoom-in to a 3.5 GHz bandwidth, resolving individual comb lines.}
    \label{fig:Optical spectrum}
\end{figure*}

\section{Quantum-enhanced gas detection}\label{sec3}
We next demonstrate EDCS for gas detection in an asymmetric scheme, in which only the signal comb is directed to the sample of interest, minimizing the risk of sample damage while maintaining a high SNR and effective technical noise suppression achieved through the use of a strong LO comb and balanced detection \cite{schumaker1984noise}. Figure~\ref{fig:Optical spectrum}a shows the optical spectrum of the LO comb with a line spacing of $f_{\rm rep}=17.565$ GHz, selected to align with the line spacing of the entangled comb while mitigating the limited resolution of the programmable optical filter. By sweeping the CW laser across 50 distinct central frequencies, we expand the spectrum to encompass 500 comb lines with a reduced spacing of 350 MHz, inferred from the RF beatnotes and depicted in Figs.~\ref{fig:Optical spectrum}b--d. The central region of the spectrum lacks comb lines due to a zero carrier-offset frequency, resulting in the absence of observable beat notes. Nonetheless, the central comb lines of the three combs are utilized to phase-lock the entangled comb displacement and to align the LO comb with the squeezed quadrature. Both locks are essential for maintaining stability in the precise measurement of the squeezed quadratures. 

\begin{figure*}[h]
\centering
  \includegraphics[width=\textwidth]{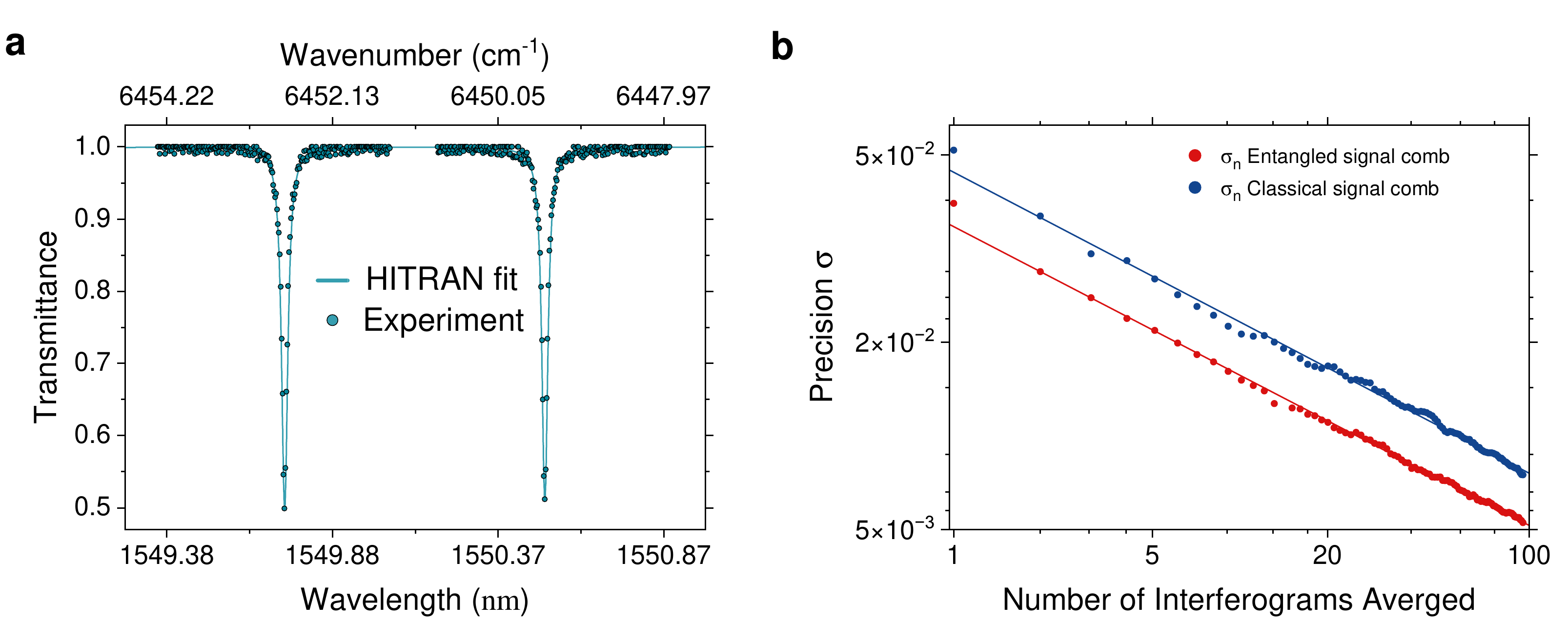}
    \caption{Measured absorption spectrum and quantum speedup. (a) Acquired absorption spectrum through EDCS and fit with HITRAN database, showcasing its accuracy. (b) Quantum speedup of EDCS over classical DCS enabled by enhanced precision below the standard quantum limit.}
    \label{fig:speed up and absorption}
\end{figure*}

In gas detection, the entangled signal comb with $\delta f_{\rm rep} = 4$ MHz passes through a gas cell containing hydrogen cyanide (HCN) at a pressure of 25 Torr, with an optical path length of 17.5 cm, yielding an absorption depth of 3 dB. To obtain the absorption spectrum of HCN, we use the comb configuration and CW laser sweep, the same as that of Fig.~\ref{fig:Optical spectrum}b. Figure \ref{fig:speed up and absorption}a shows the measured average transmittance spectrum, which is fitted to a model based on the HITRAN2020 database \cite{gordon2022hitran2020} built on the optical path length, pressure, temperature, and concentration as input parameters, showing excellent agreement between the experimental data and model.

To demonstrate the speedup enabled by EDCS, we estimate the precision of the transmittance spectrum measurement as a function of the number of averaged interferograms. The interferograms are detected using a balanced photodetector and recorded on an oscilloscope with sufficient bandwidth and a sampling rate of 100 MS/s. After filtering, a 10 µs segment of the time-domain signal is extracted, corresponding to a single interferogram with a resolution bandwidth of 100 kHz. Figure \ref{fig:speed up and absorption}b illustrates the precision as a function of the number of averaged interferograms, highlighting that EDCS achieves a quantum speedup factor of 1.7 compared to classical DCS in reaching the target precision. This advancement underscores the prospect of EDCS to overcome the fundamental trade-off between sensitivity and measurement time, paving the way for sensing applications with boosted speed and precision.

We further evaluate EDCS's performance under varying absorption depths, as illustrated in Fig.~\ref{fig:SNR}. Notably, the quantum advantage remains largely intact even with 3 dB absorption, compared to lower absorption levels of 0.1 and 0.25 dB. This robustness results from a synergistic interplay between the classical and quantum resources in our configuration. EDCS leverages the sub-SQL performance provided by the non-classical properties of TMSV states, while its distributed nature minimizes the impact of losses on quantum advantage as only a small fraction of the comb lines is expected to experience loss in a spectroscopic measurement. Crucially, a favorable unattenuated-to-attenuated ratio (UAR) of comb lines—10:1 in our setup—results in minimal degradation of sub-SQL performance. To further illustrate EDCS’s capability, we simulate the performance of a flat-top entangled comb with a maximum of 10 dB squeezing and 15 dB anti-squeezing, measured using a flat-top LO. The results, shown in Fig.~\ref{fig:SNR}c, highlight the resilience of entangled combs to loss, attributed to their high UAR of comb lines.        

\begin{figure*}[h]
\centering
  \includegraphics[width=\textwidth]{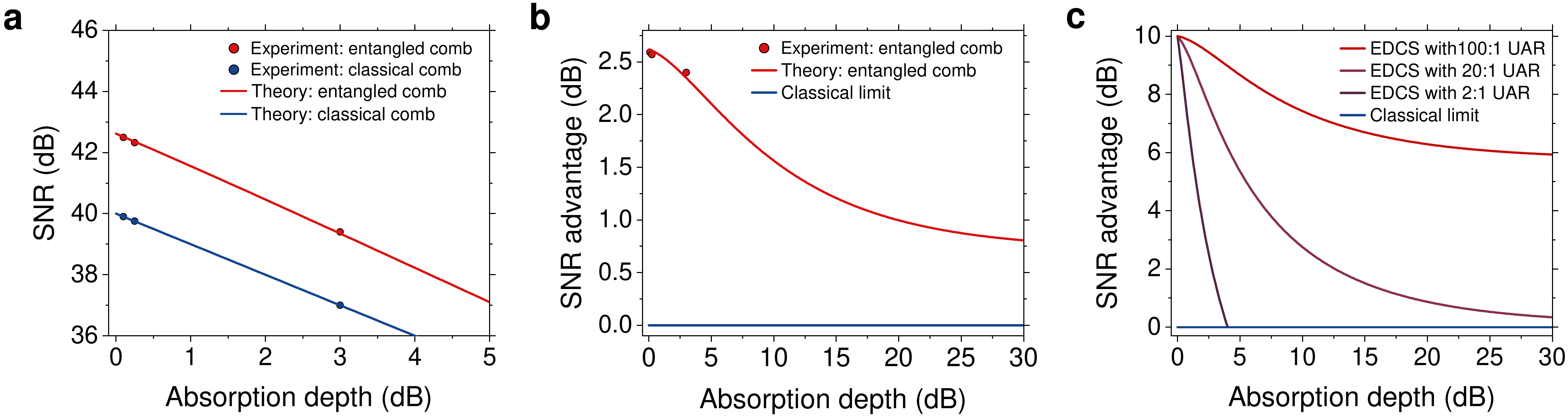}
    \caption{Robustness of EDCS's quantum advantage against sample absorption. (a) SNR of EDCS and classical DCS as a function of increasing absorption depth in gas cells. Despite the heightened attenuation at moderate depths, the quantum advantage of EDCS over classical DCS remains evident. (b) SNR advantage of EDCS over classical DCS, illustrating the moderate degradation rate of the quantum advantage with increasing absorption and its eventual saturation. (c) Simulation of a flat-top entangled comb with 10 dB maximum squeezing and 15 dB maximum anti-squeezing, measured using a flat-top LO, demonstrating the required UAR of comb lines to sustain a target quantum advantage level.}
    \label{fig:SNR}
\end{figure*}

\section{Discussion}\label{sec12}
We implement linear-absorption DCS enhanced by entangled spectral structures to achieve an enhanced SNR, enabling higher sensitivity and reduced integration times. Our experimental setup introduces an additional servo loop to lock the phase between the entangled comb and the classical signal comb, while the remaining locking mechanisms follow that of conventional TMSV sources. To address the distinct phase requirements and locking challenges in EDCS, we employ two essential techniques. The first involves EO comb generation, where a shared CW laser seeds the optical cavity as well as the classical and LO combs through EO modulation. This approach ensures intrinsic mutual coherence \cite{parriaux2020electro}, enabling sub-shot-noise detection of the entangled signal comb, shot-noise-limited detection of the classical signal comb, and precise phase locking of the LO. The second technique utilizes programmable optical filters, which provide precise control over the intensity and phase of all three combs: the entangled signal comb, the classical signal comb, and the LO comb. This control is critical for suppressing technical noise, maximizing beatnotes, and simultaneous detection of all comb pairs below the SQL. Together with precise shot-noise calibration that is essential to reliably demonstrate true sub-SQL capabilities, these methods establish the foundation for high-precision and robust measurements in EDCS.

With the current entanglement and spectral configurations, the beatings between the $n$th and $-n$th entangled signal and LO comb lines are translated into the same RF frequency at $n\delta f_{\rm rep}$. We have developed two approaches to resolve the aliasing. The first approach entails two-shot measurements, in which the positive indexed entangled comb lines are displaced by $\alpha_{S_n}$ and $-\alpha_{S_n}$ in the two consecutive measurements. Since the beatnote at $n\delta f_{\rm rep}$ in the two measurements yields $\alpha_{S_{-n}}+\alpha_{S_{n}}$ and $\alpha_{S_{-n}}-\alpha_{S_{n}}$, the sum and the difference of the two measurements unveil $2\alpha_{S_{-n}}$ and $2\alpha_{S_{n}}$ without compromising the overall SNR. In the second approach, the positive and negative indexed comb lines are displaced along two different quadratures, mathematically described as real and imaginary displacements. The beatings of the $n$th and $-n$th entangled signal and LO comb lines then yield RF signals proportional to $\cos(2\pi n\delta f_{\rm rep}t)$ and $\sin(2\pi n\delta f_{\rm rep}t)$, allowing the positive and negative indexed comb lines to be unambiguously separated by two RF mixers.

It is worth noting that the demonstrated EDCS protocol differs from that proposed in Ref.~\cite{shi2023entanglement} in both the structure of the quantum comb and the measurement scheme, allowing for the use of off-the-shelf TMSV sources while paving the way for integration into  \( \chi^{(3)} \)  complementary metal-oxide-semiconductor-compatible silicon-photonics platforms. These platforms have made significant advancements in quantum-light generation \cite{zhang2021squeezed,liu2024generation,jahanbozorgi2023generation}, closing the gap with bulk quantum sources that hold the record of the amount of measured squeezing~\cite{vahlbruch2016detection}. Moreover, these platforms offer the potential to tailor the dispersion profile~\cite{lucas2023tailoring,yu2021spontaneous} to generate ultra-broadband quantum combs and to integrate classical Kerr frequency combs in tandem with programmable optical filters \cite{moreira2016programmable} as the LO comb, opening a new avenue for the development of highly compact and versatile EDCS platforms capable of outperforming their classical counterpart \cite{dutt2018chip}.

As highlighted in \cite{xu2024near,millot2016frequency,deriushkina2022dual}, the limited spectral span—and in our case, a smaller number of comb lines—characteristic of DCS using EO modulators is effectively mitigated through the frequency tunability of the CW laser and RF synthesizers. This agility allows for precise adjustments to the center frequency and repetition frequency of the combs, providing substantial flexibility in experimental configurations. Despite these constraints, this approach delivers a high SNR within short measurement times and enables customizable spectral resolution. At present, the spectral resolution of our EDCS setup is 350 MHz shown in Fig.~\ref{fig:Optical spectrum}d. To address the remaining gaps within these intervals, the RF frequency can be swept by $\pm 175$ MHz at each laser center frequency. This two-step process guarantees complete spectral coverage while maintaining the quantum advantage, as all comb lines remain well within the 400 MHz squeezing bandwidth of the entangled comb source throughout both laser and RF sweeps. As for further expanding the spectral coverage, multiple techniques have been employed. For example, efficient nonlinear broadening can be used to achieve wider bandwidths with high resolution, as demonstrated in \cite{duran2016electro}. Additionally, cascading EO modulators, as shown in \cite{kowligy2020mid}, offers another effective method for significantly increasing spectral coverage.

EDCS's advantage over classical DCS can be further broadened via multiple routes. First, the amount of vacuum squeezing produced by our entangled-comb source is measured at $\sim 4$ dB, while sources deliver over 10 dB of squeezing have been reported~\cite{vahlbruch2016detection}. Second, the RF synthesizer employed in the present experiment exhibits a phase noise level at –75 dBc/Hz at 10 GHz frequency with a 10 kHz offset, while a standard commercially available unit offers –132 dBc/Hz at 10 GHz to further amplify the disparity in the integration times between EDCS and classical DCS. Third, mitigation of experimental imperfections, such as inaccuracies in the Waveshaper phase filter configurations, sub-unity fringe visibility (97\%) due to mode mismatch between the entangled signal and LO combs, and limited quantum efficiency of the photodiodes (88\%), will reduce mixing of the squeezed and anti-squeezed quadratures and the amount of vacuum noise mixed in, enabling an even more pronounced quantum advantage for EDCS.

Before closing, we compare our work with the two recently reported quantum DCS experiments~\cite{herman2024squeezed,yan2024quantum}. A figure of merit for DCS is the quality factor defined as ${\rm SNR}\times M/\sqrt{\tau}$, where $M = 2N$ is the total number of comb lines, SNR is defined in amplitude, and $\tau$ is the integration time~\cite{coddington2016dual}. The asymmetric DCS scheme, in which only the signal comb interrogates the sample while the other much more intense comb serves as an LO at the heterodyne detector, is known to saturate classical DCS's quality-factor limit given the power on the sample~\cite{shi2023entanglement}. Exploiting entanglement to enhance the SNR thus enables our asymmetric EDCS to surpass the fundamental limit of classical DCS. Ref.~\cite{herman2024squeezed}'s experiment, in contrast, employs a symmetric DCS scheme, in which both combs pass through the sample. The optimal quality factor for symmetric DCS subject to the same power constraint is achieved by evenly distributing the optical power over the two combs. Hence, the large imbalance between the power levels of the two combs in the experiment may render the attained quality factor inferior to the classical limit. Ref.~\cite{yan2024quantum}'s experimental DCS scheme is asymmetric based on a classical comb interrogating the sample and a squeezed LO. While such a configuration may offer some practical benefits due to the low frequency conversion efficiency, it cannot beat the fundamental limit of classical DCS~\cite{ralph2000can}.

\newpage


\bmhead{Acknowledgments}
This work is supported by Office of Naval Research Grant No. N00014-23-1-2296, National Science Foundation Grant No. 2317471 and 2326746, and University of Michigan. QZ and HS also acknowledges support from AFOSR MURI FA9550-24-1-0349 and HR001123S0052. We thank J. Wu for discussion on phase locking.

\section*{Declarations}

\bmhead{Author contributions}
AH and ZZ designed the experiment. AH performed the experiment and acquired the data. AH and SL processed the data. HS and QZ developed the theoretical model. XF advised on gas sensing. All authors analyzed the experimental results. AH and ZZ drafted the manuscript with inputs from others. ZZ conceived and supervised the project.

\section*{Appendix}
\subsection*{Experimental setup}

A detailed experimental diagram is shown in Supplementary Fig.~\ref{fig:exp scheme}. A 1550-nm fiber laser (NKT Photonics Koheras BASIK) generates 30 mW of light, which is modulated by a fiber-based EO phase modulator (PM) driven at 88 MHz to create sidebands for cavity locking via the Pound-Drever-Hall technique. The modulated light is amplified to $\sim$1.5 W by an erbium-doped fiber amplifier (EDFA) and coupled into free space. The 1550-nm light is directed to a three-mirror cavity for spatio-temporal mode cleaning (MC) and locked at its maximum transmission. The output from the MC cavity is then split into two arms: one serves as the pump for second-harmonic generation (SHG), while the other passes through a second MC cavity and is subsequently split into three arms to serve as the seed for the OPO, the classical comb, and the LO comb for heterodyne measurements.

The classical comb is generated using two fiber-based EO-PMs: one driven at 17.569 GHz to create the comb lines and another driven at 100 kHz by a lock-in amplifier to generate sidebands for locking the classical comb to the entangled comb. Prior to interfering with the entangled comb, the classical comb passes through a programmable optical filter (Waveshaper) to control the intensity and phase of each line of the entangled signal comb. The Waveshaper balances the intensities of the two comb lines forming each pair, deliberately reduces the power of all comb lines (except the central comb line used for locking) to mitigate phase noise in the RF beatnotes (Fig.~\ref{fig:RF spectrum}b,c), and optimizes the phase of each comb line to maximize RF beatnotes. Additionally, it configures the comb phase for the two-shot measurements used to resolve comb aliasing.

The SHG cavity is a semi-monolithic design incorporating a curved mirror with low reflectivity at 775 nm and high reflectivity at 1550 nm, along with a periodically-poled KTiOPO4 (PPKTP) crystal. It is locked using the 88-MHz sidebands and generates 775-nm light through second-order frequency doubling. The 775-nm light is directed to an MC cavity before being injected into the OPO cavity, which contains an identical PPKTP crystal. The OPO cavity’s curved mirror has high a reflectivity at 775 nm and a moderate reflectivity at 1550 nm. The three MC cavities are locked by tapping the reflected light from their input coupling mirrors, while the SHG cavity and the OPO’s 1550-nm and 775-nm locks are achieved by tapping their transmitted beams.

To generate an entangled comb, the weak 1550-nm seed beam is modulated using a free-space PM driven at 26 MHz, creating sidebands for parametric amplification locking. The OPO cavity is locked using a 46-MHz sideband of the 775-nm pump, which transmits through the DBS on the pump input path. When pumped with 775-nm light, the OPO generates even photon-number states via SPDC at 1550 nm. When phase-locked to achieve parametric amplification, altering the seed’s photon statistics from a coherent state to a displaced phase-squeezed state. The quantum light emitted from the OPO comprises a single-mode squeezed state at the central frequency and an entangled comb consisting of pairs of TMSV states. After its generation, the entangled comb is interfered with a classical comb on a 99/1 beam splitter, converting each TMSV pair into a displaced two-mode squeezed state. 

\subsection*{Characterization of entangled comb}
\begin{figure*}[h]
\captionsetup{name=Supplementary Fig.}
 \centering
   \includegraphics[width=\textwidth]{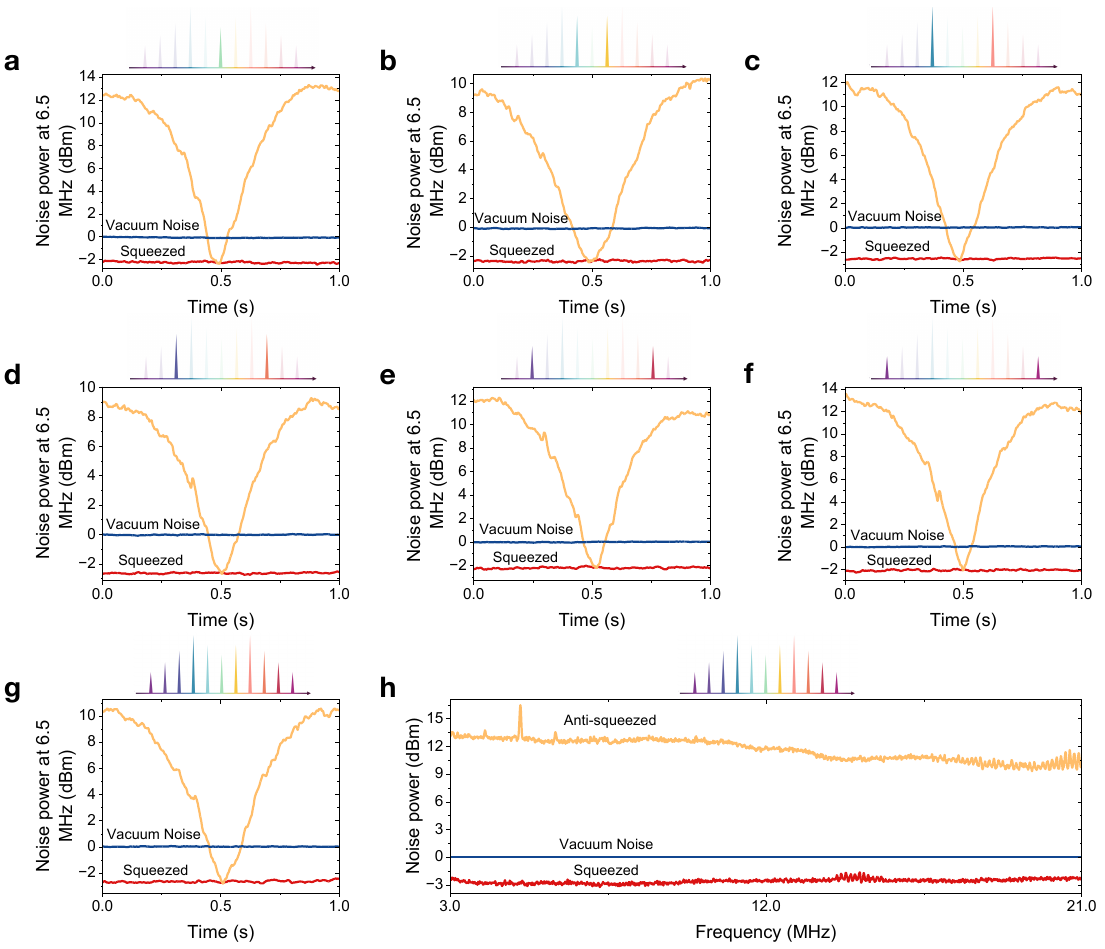}
     \caption{Zero-span measurements of the entangled comb at a frequency of 6.5 MHz, captured using an electrical spectrum analyzer.
(a) Measurement of the central comb line using LO's central comb line, while all other comb lines are off. (b) Measurement of the first entangled mode pair. (c) Measurement of the second entangled mode pair. (d) Measurement of the third entangled mode pair. (e) Measurement of the fourth entangled mode pair. (f) Measurement of the fifth entangled mode pair. (g) Measurement with all entangled come lines turned on. Measured squeezing factors range from 2.1 to 2.8 dB, and the anti-squeezing factors range from 9.3 to 13.3 dB. (h) Squeezing and anti-squeezing levels across the spectrum from 3 to 21 MHz for all comb lines measured simultaneously. The detector’s electrical noise is 18 dB below the vacuum noise. The resolution bandwidth is set to 300 kHz, and the video bandwidth to 300 Hz.}
     \label{fig:comb sqz}
 \end{figure*}

The LO comb is prepared in a manner similar to the classical comb, using a separate PM and a Waveshaper. The LO comb’s amplitudes and phases are adjusted to align with the squeezed quadratures of the TMSV states, enabling simultaneous measurement of maximum squeezing across all entangled comb lines. Without proper amplitude balancing, the LO is not matched to the signal comb, resulting in reduced observed squeezing. Furthermore, improper phase alignment results in mixing of squeezed and anti-squeezed quadratures from different modes, nullifying the overall sub-SQL capabilities of the entangled comb. Supplementary Fig.~\ref{fig:comb sqz}a-f illustrates the measured squeezed and anti-squeezed quadratures for the central comb line and five two-mode pairs. Supplementary Fig.~\ref{fig:comb sqz}g demonstrates simultaneous detection of the squeezed quadratures for all two-mode squeezed states, while Supplementary Fig.~\ref{fig:comb sqz}h demonstrates the quantum advantage over the spectrum of interest, 3-21 MHz used to obtain Fig.~\ref{fig:RF spectrum},\ref{fig:Optical spectrum} and \ref{fig:speed up and absorption}. 

In our experiment, the entangled signal comb is generated with a total power of 4 nW (Fig.~\ref{fig:RF spectrum}a), consisting of 10 signal comb lines alongside a central comb line with a power of 2 µW used to lock the phase between the entangled comb and the classical signal comb. To reduce beatnotes phase noise, the signal comb power is reduced to sub nW (Fig.~\ref{fig:RF spectrum}b,c). The entangled signal comb is measured using a LO comb with an average power of 1.8 mW per comb line. 

\subsection*{Dual-comb spectra}
The dual-comb spectra shown in Fig.~\ref{fig:RF spectrum}a-c and Figs.~\ref{fig:Optical spectrum}b-d are derived by sampling the time-domain signal of the subtracted photocurrents from the balanced receiver. The signal is sampled at a rate of 100 MHz over a duration of 0.5 seconds. The recorded data is post-processed to obtain the corresponding frequency-domain spectra. For Fig.~\ref{fig:RF spectrum}a, a resolution bandwidth of 10 kHz and 5,000 interferograms are used in the post-processing. In contrast, Figs.~\ref{fig:RF spectrum}b,c employ a resolution bandwidth of 200 Hz with 100 interferograms, while Figs.~\ref{fig:Optical spectrum}b-d use a resolution bandwidth of 100 Hz with 50 interferograms.

Finally, we note that the theoretical results shown in Fig.~\ref{fig:SNR}c can be further optimized, illustrating the additional degrees of freedom and key considerations when transitioning from DCS to EDCS. One optimization approach involves minimizing anti-squeezing by operating the OPO cavity at lower pump powers, without degrading the squeezing. This reduces the contribution of anti-squeezing to the degradation of quantum advantage observed in Fig.~\ref{fig:SNR}c. A second approach employs adaptive measurements to configure the LO in a manner that mimics the imbalance in intensity caused by absorption loss on the signal comb. This further mitigates the impact of anti-squeezing on the system’s performance. These dependencies highlight the intricate nature of EDCS and underscore its potential for significant metrological advantages.

 \begin{figure*}[h]
\captionsetup{name=Supplementary Fig.} 
    \setcounter{figure}{0}
 \centering
   \includegraphics[width=\textwidth]{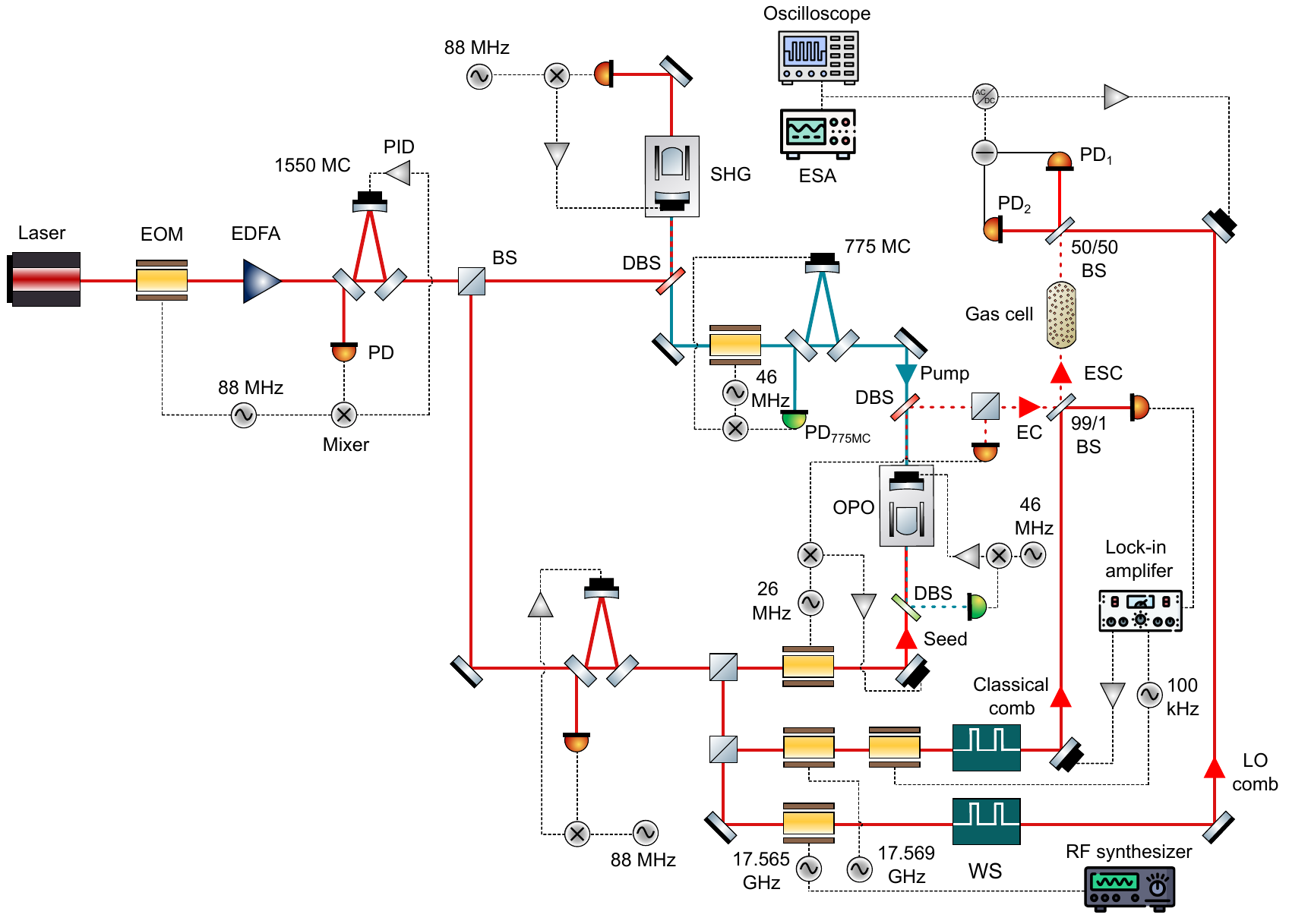}
     \caption{Schematic of the experimental setup. An entangled comb at a wavelength of 1550 nm is generated in a doubly resonant optical parametric oscillator (OPO) operated below threshold. EOM: electro-optical modulator; EDFA: erbium-doped fiber mmplifier; 1550MC: three-mirror cavity for spatio-temporal mode cleaning of 1550 nm light; PID: proportional–integral–derivative controller; BS: beam splitter; DBS: dichroic beam splitter; SHG: second harmonic generation; PD: photodiode. To generate the entangled signal comb, the entangled comb is interfered with a classical comb, generated via an EOM followed by a Waveshaper (WS), on a 99/1 BS. To lock the phase between the classical comb and entangled comb, a lock-in amplifier drives a second EOM on the classical comb arm to generate sidebands for the error signal, which is used by a PID controller to stabilize the phase. The entangled signal comb passes through a gas cell, after which it interferes with the local oscillator (LO) comb in a heterodyne configuration. The subtracted photocurrent from balanced detection is analyzed by an electronic spectrum analyzer (ESA) for real-time optimizations of the system, while an oscilloscope is used for interferogram acquisition.} 
     \label{fig:exp scheme}
 \end{figure*}
\newpage 
\pagebreak
\newpage 

\newpage 

\newpage 
\pagebreak
\newpage 
\clearpage
\bibliography{EDCS} 

\end{document}